\documentclass[aps,prl,amsmath,amssymb,floatfix,twocolumn,superscriptaddress,letterpaper]{revtex4-1}
\usepackage{multirow}
\usepackage{bbold}
\usepackage{subfigure}
\usepackage{color}
\usepackage{xcolor}
\usepackage{mathrsfs}
\usepackage{hyperref}
\usepackage[normalem]{ulem}
\usepackage{bm}

\usepackage[utf8]{inputenc}

\usepackage{amsfonts, relsize, color}
\usepackage{graphics}
\usepackage{comment}
\usepackage{graphicx}
\usepackage{subfigure}

\usepackage{ulem}

\def\ket#1{|#1\rangle }
\def\bra#1{\langle #1 |}

\begin{document}
\title{\bf Determination of spin-orbit interaction in semiconductor \\ nanostructures via non-linear transport}

\author{Renato M. A. Dantas}\email{renatomiguel.alvesdantas@unibas.ch}
\affiliation{Department of Physics, University of Basel, Klingelbergstrasse 82, 4056 Basel, Switzerland}

\author{Henry F. Legg}
\affiliation{Department of Physics, University of Basel, Klingelbergstrasse 82, 4056 Basel, Switzerland}

\author{Stefano Bosco}
\affiliation{Department of Physics, University of Basel, Klingelbergstrasse 82, 4056 Basel, Switzerland}

\author{Daniel Loss}
\affiliation{Department of Physics, University of Basel, Klingelbergstrasse 82, 4056 Basel, Switzerland}

\author{Jelena Klinovaja}\email{jelena.klinovaja@unibas.ch}
\affiliation{Department of Physics, University of Basel, Klingelbergstrasse 82, 4056 Basel, Switzerland}

\date{\today}
\begin{abstract}
We investigate non-linear transport signatures stemming from linear and cubic spin-orbit interactions in one- and two-dimensional systems. The analytical zero-temperature response to external fields is complemented by finite temperature numerical analysis, establishing a way to distinguish between linear and cubic spin-orbit interactions. We also propose a protocol to determine the relevant material parameters from transport measurements attainable in realistic conditions, illustrated by values for Ge heterostructures.
Our results establish a method for the fast benchmarking of spin-orbit properties in semiconductor nanostructures.
\end{abstract}

\maketitle

\textit{Introduction} - 
Engineering spin-orbit interactions (SOIs) in semiconductor nanostructures is a crucial challenge in several branches of physics, ranging from spintronics~\cite{RevModPhys.76.323} and topological materials~\cite{doi:10.1063/5.0055997,RevModPhys.82.3045,RevModPhys.90.015001} to quantum information processing~\cite{RevModPhys.79.1217,PhysRevA.57.120,doi:10.1146/annurev-conmatphys-030212-184248}. Notably, large values of SOIs emerge in nanowires~\cite{watzinger2018germanium,Froning2021,PhysRevResearch.3.013081,Wang2022,Camenzind2022,maurand2016cmos,PhysRevLett.120.137702} and two-dimensional heterostructures~\cite{Hendrickxsingleholespinqubit2019,hendrickx2020fast,hendrickx2020four,PhysRevLett.128.126803,Jirovec2021}, where the charge carriers are holes in the valence band rather than electrons in the conduction band~\cite{WinklerSpinOrbitCoupling2003}.
In these systems, the SOIs are also completely tunable by external electric fields~\cite{DRkloeffel1,DRkloeffel3,PhysRevB.105.075308,adelsberger2022enhanced,doi:10.1002/adma.201906523,PhysRevB.98.155319,PhysRevB.103.045305,PhysRevApplied.16.054034}, yielding sweet-spots against critical sources of noise~\cite{piot2022single,PRXQuantum.2.010348,bosco2022hole, Wang2021,PhysRevLett.127.190501} and on-demand control of the interaction between qubits and resonators~\cite{yu2022strong,DRkloeffel2,PhysRevLett.129.066801,michal2022tunable,PhysRevB.102.205412,PhysRevResearch.3.013194}.

In particular, hole gases in planar germanium (Ge) heterostructures are emerging as highly promising candidates for processing quantum information~\cite{scappucci2020germanium}.
Their large SOI enables ultrafast qubit operations at low power in a highly CMOS compatible platform~\cite{Hendrickxsingleholespinqubit2019,hendrickx2020fast,hendrickx2020four} and removes the need for additional bulky micromagnets~\cite{mi2018coherent, PhysRevX.12.021026, Watsonprogrammabletwoqubitquantum2018,TakedaResonantlyDrivenSingletTriplet2020,Yoneda2020,doi:10.1126/sciadv.abn5130,philips2022universal}, offering a clear practical advantage for scaling up the next generation of quantum processors~\cite{VandersypenInterfacingspinqubits2017,Gonzalez-Zalba2021,Xue2021}.
Remarkably, in these structures, the SOI can be designed to be linear or cubic in momentum~\cite{PhysRevLett.98.097202,PhysRevB.104.115425,PhysRevB.103.085309,PhysRevB.103.125201,wang2022modelling}, greatly impacting the response of the material to external fields~\cite{arxiv.2206.11916}. Despite its potential, an efficient and simple way to measure the SOI in these materials remains elusive.

From the early classification of solid-state systems in insulators, conductors, and semiconductors~\cite{Ashcroft} to the more recent discovery of the role of geometry and topology in non-trivial band structures~\cite{RevModPhys.82.1959,RevModPhys.82.3045,RevModPhys.90.015001,RevModPhys.93.025002}, transport experiments have been at the core of condensed matter physics, providing arguably the most practical yet insightful way to probe the physics of solid-state systems. 
While most of the seminal effects, such as the integer quantum Hall~\cite{PhysRevLett.45.494,PhysRevB.23.5632,PhysRevLett.49.405,PhysRevLett.61.2015} or the anomalous Hall~\cite{RevModPhys.82.1539,doi:10.1126/science.1234414, NatureNakatsuji15, NatPhysYasuda16,NatPhysLiu2018} effect, depend linearly on the external fields, lately, an increasing number of novel transport properties in materials with non-trivial band structure have been reported in the non-linear regime~\cite{PhysRevB.79.081406,PhysRevB.88.104412,PhysRevLett.115.216806,PhysRevLett.117.146603,NatComJuan2017,NatCommTokura18,NatureQiong19,NatMatKang19,NatComKovalev20,PhysRevB.103.L201105,PhysRevLett.117.146603,doi:10.48550,doi:10.1073/pnas.2200367119,arxiv.2205.12939,PhysRevB.106.L081127}. 
A particularly fruitful direction has been the application of dc non-linear responses, such as magnetochiral anisotropy (also known as bilinear magnetoresistance) and non-linear Hall effects, to gain insight into the electronic structure of the system~\cite{natphys.6.578,natphys.5.495,PhysRevLett.123.016801,vaz2020,nat.nano.HL,PhysRevLett.128.176602,Ywang2022-nonlinear,Wang2022a,Ywang2022-bilinear}.

In this work, we employ Boltzmann transport theory to study the response of one- (1D) and two-dimensional (2D) nanostructures with linear and cubic SOI and show that these effects leave distinct signatures in the non-linear response of the system.
Moreover, numerical analyses for realistic material parameters and small finite temperatures support the zero-temperature analytics and confirm that these signatures can be measured in state-of-the-art experiments. 
This work paves the way for a functional and time-efficient experimental characterization of SOI in these semiconductor nanostructures, enabling fast benchmarking already at the material level.

\textit{1D} -
%%%%%%%%%%%%%%%%%%%%%%%%%%%%%
%%%%%%%%%%%%%%%%%%%%%%%%%%%%%
\begin{figure*}[t!]
	\includegraphics[width=\linewidth]{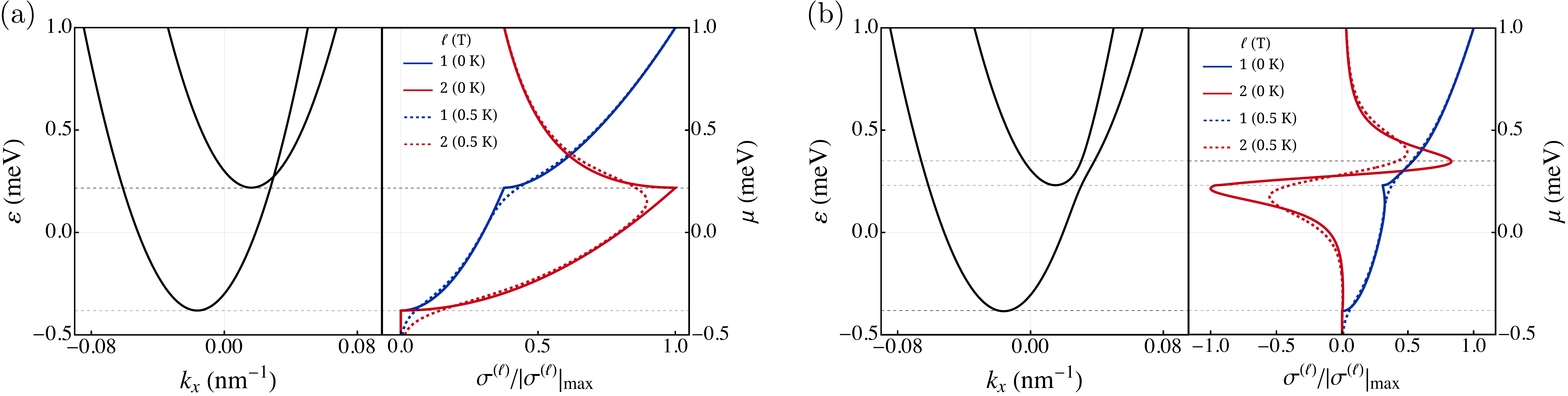}
	\caption{Band dispersions (left) and normalized $1^{\rm st}$ and $2^{\rm nd}$ order conductivities (right),  $\sigma^{(1)}$ and $\sigma^{(2)}$, {\it resp.}, for a 1D system with linear and cubic SOIs, see Eq.~(\ref{eq:H1D}), with (a) $\bm{B}=5 \, \hat{\bm{y}}$ T and (b) slightly rotated towards the SOI direction [$\bm{B}=(5 \,\hat{\bm{y}}+1 \, \hat{\bm{z}})$ T]. 
	Here, $\sigma^{({\ell})}$ for $\ell=1,2$ were numerically obtained for $T=0$ (solid lines) and $T=0.5$ K (dashed lines), for $\hbar^2 / 2m^*=310\,\rm{ meV \, nm^2}$, $\alpha =10\,\rm{meV \,nm}$, $\beta =116\,\rm{ meV \, nm^3}$, and $g_y= 2.07$~\cite{arxiv.2205.02582}.}
	\label{fig:1}
\end{figure*}
%%%%%%%%%%%%%%%%%%%%%%%%%%%%%
%%%%%%%%%%%%%%%%%%%%%%%%%%%%%
First, we consider 1D systems, e.g. nanowires, described by the effective Hamiltonian \cite{arxiv.2205.02582}
%-------------------------------------
\begin{align}
H = {}& \frac{\hbar^2 k^2_{x}}{2m^*} - \left( \alpha k_x  +\beta k^3_x \right)\sigma_y + \Delta_j \sigma_j,
\label{eq:H1D}
\end{align}
%-------------------------------------
where we assume Einstein summation convention,
$\alpha$ and $\beta$, {\it resp.}, correspond to SOI linear and cubic in momentum $k_x$, $m^*$ is the effective mass, $\sigma_i$ are the elements of the Pauli vector $\bm{\sigma}= \left( \sigma_x,\sigma_y,\sigma_z\right)$ acting in (pseudo-) spin space, and $\Delta_i= \tfrac{1}{2}\mu_{_{B}} g_i B_i$ (no summation implied) is the Zeeman field, written in terms of the components of the diagonal $g$-tensor and the external magnetic field $\bm B$~\footnote{Note that the Hamiltonian of Eq.~(\ref{eq:H1D}) is an effective description only valid for small chemicals potentials and momenta.}.
The cubic SOI $\beta$ is often neglected but it can yield significant anisotropies in the spectrum of quantum dots~\cite{Katsaros2020,arxiv.2205.02582} and, as we shall see, can be determined via transport measurements. 
The energy dispersion for such a two-band Hamiltonian is given by 
%-------------------------------------
\begin{align}
\varepsilon^{(\pm)}_{k_x}=  {}& \frac{\hbar^2 k^2_{x} }{2m^*} \pm \sqrt{\Delta^2_x+(\alpha k_x + \beta k^3_x - \Delta_y)^2+\Delta^2_z} , 
\end{align}
%-------------------------------------

The response of the system to a static external electric field, aligned with the nanowire axis ($\bm E= E \, \hat{\bm{x}}$), is obtained by solving perturbatively the Boltzmann equation within the relaxation time approximation
%-------------------------------------
\begin{equation}
f^{(s)}(\bm{k})= f_0 \big(\varepsilon^{(s)}_{\bm{k}} \big) -\tau \, \dot{\bm k}^{(s)} \cdot \nabla_{\bm k} f^{(s)}(\bm{k}),
\end{equation}
%-------------------------------------
where $f^{(s)}$ is the out-of-equilibrium distribution for the $s=\pm$ band, $f_0$ is the Fermi-Dirac distribution, and $\tau$ is the intra-band relaxation time.
In 1D, the dynamics of holes are governed solely by the electric field, i.e. $\hbar \dot{\bm k}^{(s)}= e\bm{E}$, where $e$ is the elementary charge.
The current density can be written as a power series in the electric field $E$, which at zero temperature, $T=0$, is given by
%-------------------------------------
\begin{align}
j_x{}&\!=\! \sum_l  j_x^{(l)} \!=\! \frac{e}{2\pi} \! \sum^{\infty}_{l=1} \! \left( \! \frac{ \tau e E}{\hbar} \! \right)^{\!l} \! \sum_{s,i} \! \mathrm{sgn} \bigg[v^{(s)}_{k^{\!(s)}_{\!F\!,i\!}}\bigg]  \mathcal{V}^{(s)}_{l,k^{\!(s)}_{\!F\!,i\!}} ,
\label{eq:1DJ}
\end{align}
%-------------------------------------
where $\hbar \mathcal{V}^{(s)}_{l,k_x} \! = \! \partial^l_{k_x} \varepsilon^{(s)}_{k_x}$, $\hbar v^{(s)}_{k_x} \! =\! \partial_{k_x} \varepsilon^{(s)}_{k_x}$ is the group velocity, and $k^{\!(s)}_{\!F\!,i\!}$ is the $i^{th}$ Fermi wave vector associated with the band $s$ [obtained from $\varepsilon^{(s)}_{k^{\!(s)}_{\!F\!\!}} (\bm{\Delta})=\mu$]~\cite{natphys.6.578,PhysRevLett.117.146603,nat.nano.HL}.

To gain insight into the impact of the SOI, we first consider $\bm{B}$ aligned with the spin polarization axis of the system.
Due to the SOI, the otherwise degenerate quadratic bands develop a finite spin expectation value along the $y$-direction and, in the presence of a collinear magnetic field $B_y$, become spin-split [$\varepsilon^{(s)}_{k_x} \! =\! \tfrac{\hbar^2 k^2_{x} }{2m^*} + s (\Delta_y \!- \alpha k_x -\beta k^3_x)$].
In this case, the only non-zero contribution in Eq.~(\ref{eq:1DJ}), aside from the linear term, which is dominated by the kinetic term and hence less useful to extract information about SOI, is given by the  term quadratic in $E$, 
%-------------------------------------
\begin{align}
j^{(2)}_x{}&=  - \frac{3 e^3 \tau^2 E^2 \beta}{\pi \hbar^3} \sum_{s=\pm} s \left[k^{(s)}_{F,R} -k^{(s)}_{F,L}\right],
\label{eq:1dcSOI}
\end{align}
%-------------------------------------
where $k^{(s)}_{F,R(L)}$ is the positive (negative) Fermi wave vector associated with the band $s$.
Note that $j^{(2)}_x$ is directly proportional to the cubic SOI coupling $\beta$, providing a way to directly determine the presence of this effect from the second-order conductivity $\sigma^{(2)}$, defined by $j_x = \sigma^{(1)}E+\sigma^{(2)}E^2+\mathcal{O}(E^3)$.
Alternatively, non-linear currents can be induced in systems lacking cubic SOI by choosing ${\bf B}$ such that $\bm{\Delta}\times \hat{\bm{y}} \neq 0$  (Fig.~\ref{fig:1})~\cite{arxiv.2205.12939}.

\textit{2D} - 
We now consider the effect of linear and cubic SOI in a 2D system described by the effective Hamiltonian
%-------------------------------------
\begin{align}
H = {}& \frac{\hbar^2 k_{+}k_{-}}{2m^*} +  \alpha \left( \bm{\sigma}\times \bm{k}\right) \cdot \hat{\bm{z}} + i \beta_1 \left(k^3_{-} \sigma_{+} \! - k^3_{+} \sigma_{-}\right) \nonumber\\
{}& + i \beta_2 \left( k_{-} k^2_{+} \sigma_{+} -k_{+} k^2_{-} \sigma_{-}\right) + \sigma_i \,  \Delta_{i},
\label{eq:H2D}
\end{align}
%-------------------------------------
where $k_{\pm}= k_{x} \pm i k_{y}$ and $ \sigma_{\pm} = ( \sigma_{x} \pm i \sigma_{y})/2$. Here, $\alpha$, $\beta_1$, and $\beta_2$ are the linear, isotropic cubic, and anisotropic cubic SOI constants, {\it resp.} \cite{PhysRevLett.95.076805,PhysRevLett.98.097202,PhysRevB.95.075305,PhysRevB.95.085431,PhysRevB.103.125201,arxiv.2209.12745}.
The energy dispersion reads
%-------------------------------------
\begin{align}
\varepsilon^{(\pm)}_{\bm{k}} \!=\! \frac{\hbar^2 \left( k^2_{x}+k^2_{y} \right) }{2m^*}  \pm  \sqrt{n^2_x \! + \!n^2_y \!+ \! \Delta_z^2},
\end{align}
%-------------------------------------
where
%-------------------------------------
\begin{align}
{}&n_x \!= \! \alpha  k_y \!+ \tfrac{i}{2} \! \left[\beta_1(k^3_{-} \! - \! k^3_{+} \! ) \!-\! \beta_2 ( k^2_{-}k_{+} \! - \! k^2_{+}k_{-} )\right] \!+\! \Delta_x, \nonumber \\
{}&n_y \!= \! -\alpha k_x \! - \! \tfrac{1}{2} \! \left[ \beta_1 (k^3_{-} \! + \! k^3_{+} \!) \!+ \! \beta_ 2 (k^2_{-}k_{+} \! + \! k^2_{+}k_{-} \!) \right] \! + \! \Delta_y. \nonumber
\end{align}
%-------------------------------------
To set the  parameter values, hereafter we consider  Ge [100] and [110] for realistic experimental conditions. 
While both systems have similar effective masses and cubic SOIs, assumed as $\hbar^2/2m^*= 620\,\rm{ meV \, nm^2}$, $\beta_1 =190\,\rm{ meV \, nm^3}$ and $\beta_2 =23.75\,\rm{ meV \, nm^3}$, they differ in the linear SOI, which is absent for Ge [100] and takes the value $\alpha =1.5 \,\rm{meV \, nm}$ for Ge [110]~\cite{PhysRevB.103.085309}. Further, $g_{x/y}= 0.207$ for Ge [100], while for Ge [110], $g_{x/y}= 1.244$.

%%%%%%%%%%%%%%%%%%%%%%%%%%%%%
%%%%%%%%%%%%%%%%%%%%%%%%%%%%%
\begin{figure*}[t!]
	\includegraphics[width=\linewidth]{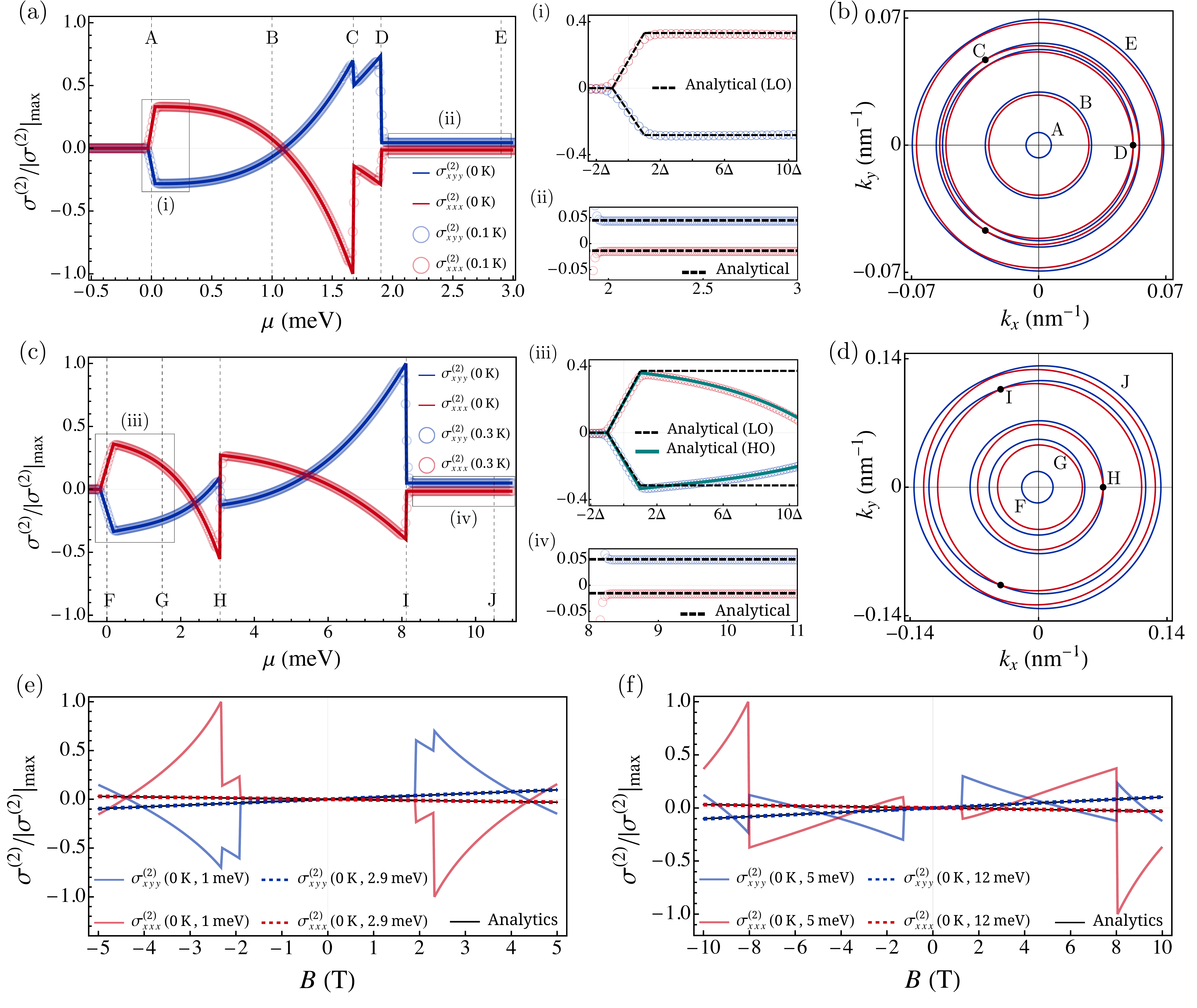}
	\caption{Normalized $2^{\rm nd}$ order longitudinal $\sigma^{(2)}_{xxx}$ and transversal $\sigma^{(2)}_{xyy}$ conductivities, as function of  Fermi energy $\mu$, for planar Ge with (a) only cubic SOI, ($\beta_1,\beta_2)=(190,23.75)\,   \rm{meV nm^3}$,  and (c) both linear ($\alpha=1.5\,\rm{meV nm}$) and cubic SOI, with $B=5$ T aligned in $y$-direction. Solid lines (circles) result from numerical evaluation of Eq.~(\ref{eq:2DJ}) at $T=0$ ($T>0$).
	The comparison between the  numerical results for $T>0$ presented in (a) [(c)] and the $T=0$ analytical expressions of Eqs.~(\ref{eq:socZbSOI}-\ref{eq:socZbSOI2}) and Eqs.~(\ref{eq:socSOIbZl1}-\ref{eq:socSOIbZ}) is shown in (i) and (ii) [(iii) and (iv)], {\it resp}. 
	In (iii) we compare results including higher order (HO) terms to the linear order (LO) analytical results from Eqs.~(\ref{eq:socZbSOI}-\ref{eq:socZbSOI2}).
	(b) [(d)] Fermi contours for  representative  $\mu$'s marked in (a) [(c)]. Normalized  $\sigma^{(2)}_{xxx}$ and $\sigma^{(2)}_{xyy}$ as  function of  $\bm{B}= B \hat{\bm{y}}$, for two representative  $\mu$'s in (e) Ge[100] and (f) Ge[110]. Colored solid and dashed lines represent results obtained from numerical evaluation of Eq.~(\ref{eq:2DJ}) at $T=0$ , while black solid  lines 
	stem from the corresponding analytics, see Eqs.~(\ref{eq:socSOIbZl1}-\ref{eq:socSOIbZ}). 
	Assuming $\tau=10$ ps~\cite{doi:10.1063/5.0083161}, we have (a) $|\sigma^{(2)}_{xxx}|_{\rm{max}} \approx 1.67$ nA m V$^{-2}$, (b) $|\sigma^{(2)}_{xyy}|_{\rm{max}} \approx 8.97$ nA m V$^{-2}$, (e) $|\sigma^{(2)}_{xxx} |_{\rm{max}}\approx 0.77$ nA m V$^{-2}$, and (f) $|\sigma^{(2)}_{xxx} |_{\rm{max}} \approx 8.76$ nA m V$^{-2}$.}
	\label{fig:Fig3}
\end{figure*}
%%%%%%%%%%%%%%%%%%%%%%%%%%%%%
%%%%%%%%%%%%%%%%%%%%%%%%%%%%%
Compared to 1D, the dynamics of the hole quasiparticles is not only enriched by orbital effects for out-of-plane magnetic fields but also by the band geometry, captured by the equations of motion
%-------------------------------------
\begin{align}
\! \dot{\bm r}^{(s)} {}&\!= \! \bm{\tilde{v}}^{(s)}_{\bm{k}} \! - \! \dot{\bm{k}}^{(s)} \! \times \bm{\Omega}^{(s)}_{\bm{k}} \! , \quad \hbar \dot{\bm k}^{(s)} \!= \! e \bm{E}  -  e \dot{\bm{r}}^{(s)} \! \times \! \bm{B} ,
\end{align}
%-------------------------------------
where $\bm{\Omega}^{(s)}_{\bm{k}} \! =\!i \bra{\nabla_{\bm{k}} u^{\!(s)}_{\bm{k}}} \! \times \! \ket{\nabla_{\bm{k}} u^{\!(s)}_{\bm{k}}}$ is the Berry curvature, with $H \ket{ u^{\!(s)}_{\bm{k}}} \! = \! \varepsilon^{(s)}_{\bm{k}} \ket{ u^{\!(s)}_{\bm{k}}}$, and $\bm{\tilde{v}}^{(s)}_{\bm{k}} \!= \! \bm{v}^{(s)}_{\bm{k}} \! - \! \hbar^{-1} \nabla_{\! \bm k}( \bm{m}^{(s)}_{\bm{k}} \! \cdot \! \bm{B})$ is the group velocity for the dispersion modified by the orbital magnetic moment of the wave packet, $\bm{m}^{(s)}_{\bm{k}} \!=\! \tfrac{i e }{2 \hbar} \bra{\nabla_{\bm{k}} u^{\!(s)}_{\bm{k}}} \! \times \![ H \!-  \varepsilon^{(s)}_{\bm{k}} \! |_{\bm{B}=0} ]\ket{\nabla_{\bm{k}} u^{\!(s)}_{\bm{k}}}$~\cite{RevModPhys.82.1959}.
Note that, because we consider holes, the quasimomentum couples with opposite sign to the electric field when compared to the case of electrons.
Since for 2D systems both $\bm{\Omega}^{(s)}_{\bm{k}}$ and $\bm{m}^{(s)}_{\bm{k}}$ are perpendicular to the plane, we can greatly simplify the problem by restricting to in-plane fields, $\bm{B} \! \perp \! \hat{\bm{z}}$. This suffices to determine both SOI and in-plane $g$-factors. This simplification renders the physics similar to 1D, allowing the result from Eq.~(\ref{eq:1DJ}) to be generalized, such that the $l^{th}$-order contribution to the current becomes 
%-------------------------------------
\begin{align}
j^{(l)}_{q}&=\frac{ e^{l+1} \tau^l}{(2\pi)^2\hbar^l}  E_{i_1} \! \cdots E_{i_l} \! \sum_{s=\pm} \oint_{\rm{FS}} \! \hat{v}^{(s)}_{i_{l}} \mathcal{V}^{(s)}_{i_{l-1} \cdots i_1  q } ds,
\label{eq:2DJ}
\end{align}
%-------------------------------------
where $\oint_{\rm{FS}}$ denotes the line integral over the Fermi surface (FS), $\hat{v}^{(s)}_{i_{l}}=v^{(s)}_{i_{l}}/|\bm{v}^{(s)} |$, and $\hbar \mathcal{V}^{(s)}_{i_1 \dots i_l} = \partial_{k_{i_1}} \cdots \partial_{k_{i_l}} \varepsilon^{(s)}_{\bm{k}}$.
This expression can be used to numerically access the response at $T=0$.
Henceforth, we focus on the lowest order non-linear response impacted by the SOI and study the $2^{\rm nd}$ order conductivities $\sigma^{(2)}_{iln}$, defined by $j_i= \sigma_{il} E_{l} + \sigma^{(2)}_{iln} E_{l} E_{n}+ \mathcal{O}(E^3)$. Importantly, second order conductivities correspond to a rectification of current and are genereally only allowed in the presence of both broken inversion symmetry and time reversal symmetry (see below), this makes them highly sensitive to the presence of SOI \cite{natphys.6.578,natphys.5.495,PhysRevLett.123.016801,nat.nano.HL,PhysRevLett.128.176602}. We use Eq.~(\ref{eq:2DJ}) to compute longitudinal, $\sigma^{(2)}_{xxx}$, and tranversal, $\sigma^{(2)}_{xyy}$, conductivities for Ge[100] and [110] for $B_y=5$ T, see Figs.~\ref{fig:Fig3}(a,c).

Since for in-plane magnetic fields linear and non-linear conductivities are solely connected to the energy dispersion [Eq.~(\ref{eq:2DJ})], these quantities reflect the competition between the kinetic, Zeeman, and SOI energies. 
We identify three different regimes from the conductivities [Fig.~\ref{fig:Fig3}(a) and (c)] which can be used to fully determine the SOI and in-plane $g$-factors.
At small chemical potentials $\mu$, the energy associated with SOI sets the smallest energy scale in the problem, allowing for the analytical computation of the conductivities. To linear order in SOI and for a generic in-plane magnetic field, we find 
%-------------------------------------
\begin{align}
\! \sigma^{(2)}_{x x x} \! &=\! - 3 \mathcal{B}  \left( \beta_1 \! + \! \beta_2\right) \! \hat{\Delta}_y , && \sigma^{(2)}_{y x x} \! =\! \mathcal{B}  \left( 3 \beta_1 \! - \! \beta_2\right) \!\hat{\Delta}_x  , \label{eq:socZbSOI} \\
\! \sigma^{(2)}_{y y y} \!&= \! - 3 \mathcal{B} \left( \beta_1 \!+\! \beta_2\right) \! \hat{\Delta}_x ,&& \sigma^{(2)}_{x y y} \! =\! \mathcal{B} \left( 3 \beta_1 \! - \! \beta_2\right) \! \hat{\Delta}_y , 
\end{align}
%-------------------------------------
and $\! \sigma^{(2)}_{y x x} \! = \sigma^{(2)}_{x y x} = \sigma^{(2)}_{x x y}$, $\sigma^{(2)}_{x y y}\!=\sigma^{(2)}_{y x y} =\sigma^{(2)}_{y y x}$, with 
%-------------------------------------
\begin{align}
\mathcal{B}= \gamma \big[ \left( \mu -\Delta \right) \Theta\left(\mu-\Delta \right) -\left( \mu + \Delta \right) \Theta\left(\mu + \Delta \right)  \big],
\label{eq:socZbSOI2}
\end{align}
%-------------------------------------
$\gamma=\tfrac{\tau^2 e^3 m^*}{\pi \hbar^5}$, $\hat{\Delta}_i=\Delta_i/\Delta$ and $\Delta^2=\Delta_i \Delta_i$.
The comparison between these expressions and the numerical results obtained with Eq.~(\ref{eq:2DJ}) are presented in Figs.~\ref{fig:Fig3}(a.i) and~\ref{fig:Fig3}(c.iii).
As $\mu$ increases, the SOI becomes more significant and higher-order terms in the SOI need to be included [see Fig.~\ref{fig:Fig3}(c.iii) for comparison of linear and higher-order analytical results]. Although full analytical treatment is possible, the expressions are too lengthy to be shown here.

The second regime emerges when the SOI becomes comparable to the Zeeman energy $\Delta_i$. This regime is delimited by the Fermi energies where the two Fermi contours touch each other [see contours C and D in Fig.~\ref{fig:Fig3}(b) and contours H and I in Fig.~\ref{fig:Fig3}(d)], leading to a discontinuity in the $2^{\rm nd}$ order 
conductivities [Figs.~\ref{fig:Fig3}(a) and~\ref{fig:Fig3}(c)]. For $\mu$ beyond this point, the response can be computed perturbatively in $\Delta_i$,
%-------------------------------------
\begin{align}
\! \sigma^{(2)}_{x x x} \! {}&= \! - \frac{3 \gamma }{4}  (\beta_1 \!  - \! 5 \beta _2 ) \Delta_y,  \quad \! \sigma^{(2)}_{y x x} \! =\! \frac{\gamma}{4} ( 3 \beta_1 \!  + \! 5 \beta _2 ) \Delta_x, \label{eq:socSOIbZl1}\\
\! \sigma^{(2)}_{y y y} \! {}&= \!  - \frac{3 \gamma }{4}  (\beta_1 \!  - \! 5 \beta _2 ) \Delta_x,  \quad \! \sigma^{(2)}_{x y y} \!=\! \frac{\gamma}{4} ( 3 \beta_1 \!  + \! 5 \beta _2 ) \Delta_y ,
\label{eq:socSOIbZ}
\end{align}
%-------------------------------------
$\sigma^{(2)}_{y x x} = \sigma^{(2)}_{x y x} = \sigma^{(2)}_{x x y} $ and $\sigma^{(2)}_{x y y}=\sigma^{(2)}_{y x y}=\sigma^{(2)}_{y y x}$. 
Here, we have neglected the linear SOI and assume that $\beta_1>\beta_2 \ge 0$.
Yet, these expressions describe extremely well the numerical results based on Eq.~(\ref{eq:2DJ}) even for systems with small linear SOI, see Figs.~\ref{fig:Fig3}(a.ii) and~\ref{fig:Fig3}(c.iv).

It is important to note that the expressions in Eqs.~(\ref{eq:socZbSOI}-\ref{eq:socSOIbZ}) do not depend on the linear SOI.
Nevertheless, the amplitude of the linear SOI can be inferred from the position of the discontinuities in the second-order conductivities, which for 
${\bf B}$ in $y$-direction are given by \footnote{Note that without linear SOI, corresponding to Ge [100], the discontinuities in $\sigma^{(2)}$ occur at $\mu_{_{\rm{C}}}= \tfrac{\hbar^2 |\Delta_y|^{2/3}}{2 m^* |\beta_1+\beta_2|^{2/3}} $ and $\mu_{_{\rm{D}}}= \tfrac{\hbar^2 |\beta_1\Delta_y|}{2 m^* |\beta_1(\beta_1-\beta_2)|^{2/3}| (\beta_1+ \beta_2) \Delta_y|^{1/3}}$.}
%-------------------------------------
\begin{equation}
\mu_{_{\rm{H}}} \! = \! \frac{\hbar^2}{2m^*} \! \left(\! \frac{\alpha}{\mathcal{X}} \! - \! \frac{\mathcal{X}}{3\beta_{+}} \! \right)^{\!2\!}\!, \, \mu_{_{\rm{I}}} \!=\! \frac{\hbar^2 \! \left(2 \alpha \!+\!4 \alpha^2 \mathcal{Y}^{-1}\! + \! \mathcal{Y} \right)}{12 \, m^* \beta_{+}},
\label{eq:pk1}
\end{equation}
%-------------------------------------
with
%-------------------------------------
\begin{align}
\mathcal{Y}={}&\! \frac{\sqrt[3]{2/3}}{\beta_1 \beta^2_{-}} \! \bigg[ \! \sqrt{\! \beta^3_1\beta_{-}^4 \! \left(12 \alpha ^3 \beta_{-}^2 \! + \! 81 \beta_1 \beta_{+}^2 \Delta_y^2 \right)} \! - \! 9 \beta^2_1 \beta_{-}^2 \beta_{+} \Delta_y \! \bigg]^{\!\frac{2}{3}\!}\!, \nonumber \\
\mathcal{X}={}&\! \sqrt[3]{\frac{3}{2}} \bigg[ \! \sqrt{\beta_{+}^3 \! \left(\! 12 \alpha^3+81\beta_{+} \Delta^2_y\right)} +9\Delta_y \beta_{+}^2\bigg]^{\! \frac{1}{3}}\!,
\label{eq:pk2}
\end{align}
%-------------------------------------
and $\beta_{\pm}=\beta_1\pm\beta_2$. This information can also be inferred by varying $B$ instead of $\mu$, see Figs.~\ref{fig:Fig3}(e) and~\ref{fig:Fig3}(f).

\begin{figure}[h!]
	\includegraphics[width=\linewidth]{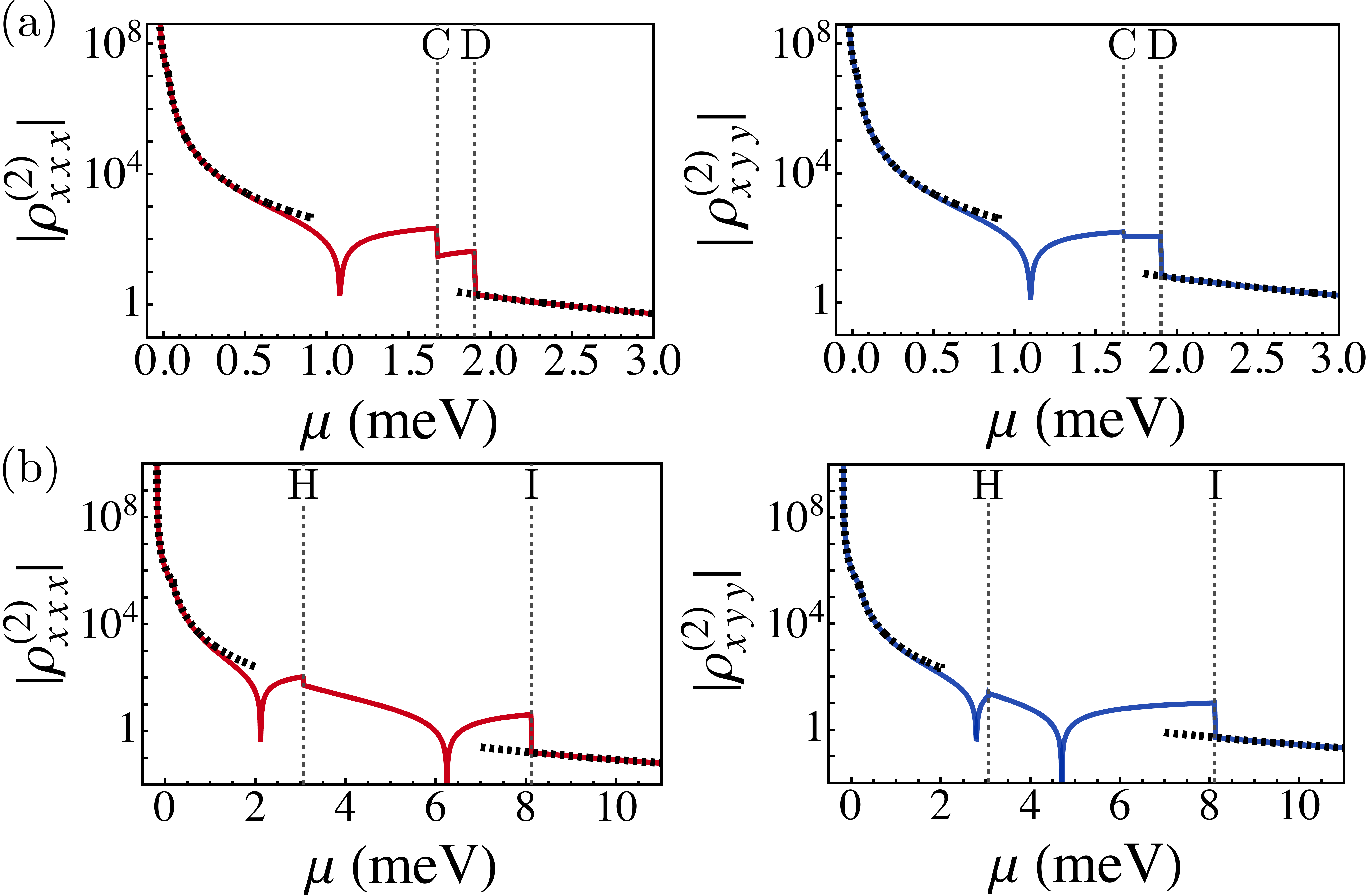}
	\caption{
	Longitudinal $\rho^{(2)}_{xxx}$ (left panel) and transversal $\rho^{(2)}_{xyy}$ (right panel) resistivities, in units of m$\Omega\,$mm$\,\rm{A}^{-1}$, as function 
	of $\mu$, for planar Ge with (a) only cubic SOI, ($\beta_1,\beta_2)=(190,23.75)\,   \rm{meV nm^3}$,  and (b) both linear ($\alpha=1.5\,\rm{meV nm}$) and cubic SOI, with $B=5$ T aligned in $y$-direction. 
	We assume $\tau=10$ ps~\cite{doi:10.1063/5.0083161}.
	Solid lines result from the numerical evaluation of Eq.~(\ref{eq:2DJ}) at $T=0$, while dashed black  lines stem from the analytical results in Eqs.~(\ref{eq:socZbSOI}-\ref{eq:socSOIbZ}) and~(\ref{eq:C2R}-\ref{eq:FOC}) for $T=0$. The values of $\mu$ at which $\sigma^{(2)}$ has discontinuities (due to the touching of the Fermi contours) are marked with vertical dashed lines as in Fig.~\ref{fig:Fig3}.}
	\label{fig:Fig4}
\end{figure}

\textit{Experimental realization} - The conductivities $\sigma^{(2)}_{iln}$ are straightforwardly related to the experimentally measured second-order resistivities $\rho^{(2)}_{iln}$, defined by $E_i=\rho_{il} j_l \approx  \rho^{(1)}_{il} j_l+ \rho^{(2)}_{iln} \, j_l j_n $, via \cite{PhysRevB.105.045421}
%-------------------------------------
\begin{align}
\rho^{(1)}_{i l} =\sigma^{-1}_{i l} ,\quad \rho^{(2)}_{iln} =- \rho^{(1)}_{i p} \sigma^{(2)}_{p q r \vphantom{l}} \rho^{(1)}_{q l}  \rho^{(1)}_{r n \vphantom{l}}.
\label{eq:C2R}
\end{align}
%-------------------------------------
Here, $\sigma^{-1}$ represents the inverse matrix of the first-order conductivity tensor, which is well described by
%-------------------------------------
\begin{align}
\sigma_{il} \! = \! \frac{\tau e^2}{2 \pi\hbar^2}\left[ \left(\mu \! - \! \Delta \right) \! \Theta  \! \left(\mu \! - \! \Delta \right) \! + \! \left( \mu \! + \! \Delta \right) \! \Theta  \! \left(\mu+\Delta \right) \right] \! \delta_{i l}.
\label{eq:FOC}
\end{align}
%-------------------------------------
In Fig.~\ref{fig:Fig4} we show the evaluation of $\rho^{(2)}_{iln}$ for Ge[100] and Ge[110].
We note that $\rho^{(2)}_{iln}$ can be obtained from standard transport experiments by measuring the second-harmonic component of the voltage induced by an ac \footnote{Here, the frequency $f$ of the ac current is chosen such that $f \tau \ll 1$. This way, the high-harmonics probe only the dc properties of the system and contain no information about the dynamical response of the system to the oscillatory driving perturbation~\cite{NatComKovalev20,PhysRevB.103.L201105,PhysRevLett.128.176602,PhysRevB.106.L081127}.} current~\cite{natphys.6.578,NatCommTokura18,natphys.5.495,PhysRevLett.123.016801,nat.nano.HL}.

%%%%%%%%%%%%%%%%%%%%%%%%%%%%%
%%%%%%%%%%%%%%%%%%%%%%%%%%%%%
%%%%%%%%%%%%%%%%%%%%%%%%%%%%%
%%%%%%%%%%%%%%%%%%%%%%%%%%%%%

Finally, we address the extraction of the SOI couplings and in-plane $g$-factors for a 2D system.
As shown in Fig.~\ref{fig:Fig4}, Eqs.~(\ref{eq:socZbSOI}-\ref{eq:socSOIbZ}) and~(\ref{eq:C2R}-\ref{eq:FOC}) provide a good fit to the numerical results for $\mu \sim \Delta$ and $\mu \gg \Delta$.
Consequently, cubic SOI can be inferred from the measurement of the first- and second-order resistivity (or resistance) tensor in any of these regimes. Note, however, that, although the non-linear resistivity is smaller for $\mu \gg \Delta$, this regime has the advantage of not requiring strong in-plane magnetic fields and being more robust to temperature fluctuations. Moreover, kinetic theory is most adequate to model the transport properties of the system in this regime, as the effect of disorder is less pronounced. Lastly, we remark again that the linear SOI coupling and the in-plane $g$-factors can then be extracted from the position of the discontinuitites in the resistivity, using Eqs.~(\ref{eq:pk1}-\ref{eq:pk2}).

\textit{Conclusion} - We discussed the effect of linear and cubic SOIs on the transport properties of 1D and 2D systems with large SOI and how non-linear responses can be used to characterize this effect with standard transport measurements. 
In nanowires, we demonstrated that the cubic SOI induces a non-linear response current quadratic in the electric field [Eq.~(\ref{eq:1dcSOI})] when the system is subjected to an external magnetic field aligned with the spin polarization axis [Fig.~\ref{fig:1}(a)]. Moreover, the linear SOI was shown to contribute to this non-linear current for noncollinear configurations of magnetic field and SOI direction, substantiating the use of the second-order conductivity to extract both $\alpha$ and $\beta$ [Fig.~\ref{fig:1}(b)]. 
Linear and cubic SOI were also proven to leave their imprint on the non-linear response of 2D systems in the presence of in-plane magnetic fields. More specifically, we showed that both longitudinal [$\sigma^{(2)}_{xxx}$ and $\sigma^{(2)}_{yyy}$] and transversal [$\sigma^{(2)}_{xyy}$ and $\sigma^{(2)}_{yxx}$] second-order conductivities can be used to determined not only the SOI couplings, but also the in-plane $g$-factors of the system [Eqs.~(\ref{eq:socZbSOI}-\ref{eq:socSOIbZ})].
 Numerical results for finite temperatures and realistic material parameters substantiate our zero-temperature analytical findings, establishing an operative characterization of the SOI and in-plane $g$-factors from measurements of non-linear resistivities in simple transport experiments.

\begin{acknowledgments}
\textit{Acknowledgements} - This work was supported by the Georg H. Endress Foundation and as a part of NCCR SPIN funded by the Swiss National Science Foundation (grant no. 51NF40-180604). This project received funding from the European Union's Horizon 2020 research and innovation program (ERC Starting Grant, grant agreement No. 757725).
\end{acknowledgments}

\bibliography{SOIC}
	
\end{document}